\documentclass[amsmath,twocolumn,aps,prl]{revtex4}
\usepackage{graphicx}
\usepackage{graphics}
\usepackage{dcolumn}
\usepackage{mathptmx}  

\begin{document}

\title{A Nanoscale Experiment Measuring Gravity's Role in \\ Breaking the Unitarity of Quantum Dynamics}

\author{Jasper  van Wezel}
\affiliation{
Cavendish Laboratory, University of Cambridge, Madingley Road, Cambridge, UK.\footnote{Current address: Materials Science Division, Argonne National Laboratory, Argonne, IL 60439, USA.}  }
\author{Tjerk H. Oosterkamp}
\affiliation{Leiden Institute of Physics, Niels Bohrweg 2, 2333 CA  Leiden, The Netherlands.}

\begin{abstract}
Modern, state of the art nanomechanical devices are capable of creating spatial superpositions that are massive enough to begin to experimentally access the quantum to classical crossover, and thus force us to consider the possible ways in which the usual quantum dynamics may be affected. One recent theoretical proposal describes the crossover from unitary quantum mechanics to classical dynamics as a form of spontaneous symmetry breaking.  Here, we propose a specific experimental setup capable of identifying the source of unitarity breaking in such a mechanism. The experiment is aimed specifically at clarifying the role played by gravity, and distinguishes the resulting dynamics from that suggested by alternative scenarios for the quantum to classical crossover. We give both a theoretical description of the expected dynamics, and a discussion of the involved experimental parameter values and the proposed experimental protocol.
\end{abstract}

\maketitle

\section{Introduction}
Experimentalists are pushing to cool cantilevers with low intrinsic damping to ever lower temperatures, both in the context of magnetic resonance force microscopy (MRFM) \cite{Rugar}, and in the quest to cool a mechanical resonator to its zero point motion \cite{Cleland,Roukes,Schwab} and eventually bring it into a quantum mechanical superposition \cite{Schwab,Bouwmeester,Roukes2,Schwab2}. One such experiment has recently succeeded in cooling a resonator to its quantum mechanical ground state and has demonstrated the controllable creation of single quantum excitations \cite{Cleland}. Several other feasible experiments have been proposed in which a mechanical resonator is coupled to a quantum system in such a way that, with some experimental progress, a superposition of spatially separated states of the resonator may be detectable \cite{Bouwmeester,Schwab2}. 
With the advent of such experiments, involving both quantum mechanics and mesoscopic objects, the problem of reconciling quantum mechanics with classical physics is no longer a purely theoretical endeavor, but rather becomes an experimental necessity. 

Many theoretical proposals exist for how to explain the apparent absence of (spatial) quantum superpositons of macroscopic objects in the classical world. The most well known of these include the idea that environment induced decoherence hides the superposed states from our view \cite{decoherence}; the idea that our personal participation in superpositions implies an inability to observe alternative branches of the superposed state of the universe \cite{manyworlds}; and the idea that corrections to Schr\"odinger's equation become important at a scale intermediate between that of microscopic particles and macroscopic objects \cite{GRW,CSL0,CSL,CSL2}. Recently, it has been pointed out by several authors that gravity may play an important role in scenarios of the latter kind \cite{CSL0,Diosi,Penrose,vWezel:PhilMag}. One particular very recent proposal posits that the transition between quantum dynamics at the microscopic scale and classical physics at the macroscopic scale may be described as a form of spontaneous symmetry breaking, in direct analogy to the breaking of translational symmetry in macroscopic crystals, rotational symmetry in macroscopic magnets, and so on \cite{vWezel:Review}. The particular symmetry being broken in this case is the unitarity of quantum mechanical time evolution, and the necessary symmetry breaking field may be supplied by the subtle influence of gravity at mesoscopic length scales.

Here, we present a specific experiment which we believe is capable of producing a superposition that is massive enough to test the theory of spontaneously broken unitarity, while allowing enough experimental control to also differentiate its predictions from those of alternative scenarios. The suggested setup is based on realistic estimates of experimental parameters which may be obtained in state of the art MRFM experiments. The classical object being forced into a superposition is the micromechanical resonator which forms the detector arm of a typical MRFM experiment. This setup differs from similar proposals in the literature \cite{Bouwmeester,Schwab2} because of the entanglement of the resonator with a nearby microscopic spin state, which allows a definitive differentiation between the effects of decoherence, spontaneously broken unitarity, and alternative scenarios, using the specific experimental protocol described here.

In the following, we first describe the proposed experimental setup, and discuss the parameters that will have to be fulfilled in order for the experiment to be successful. We then present numerical simulations of the expected time evolution of this setup in the context of spontaneously broken unitarity, which explicitly show how gravity may influence the quantum dynamics of the resonator. We also show how a pointer basis (which defines the possible outcomes of a measurement on a quantum system) and Born's rule (which deals with the probability that such a measurement yields a given result), are automatically recovered from the time evolution of the resonator in this scenario. Finally, we turn to a detailed description of the experimental protocol required to distinguish the specific time evolution discussed here from both more common disturbances to quantum dynamics, such as decoherence, and from alternative theoretical scenarios.

\section{The Experimental Setup}
\label{section2}
\begin{figure}
      \begin{center}
      \includegraphics[width=0.8\columnwidth]{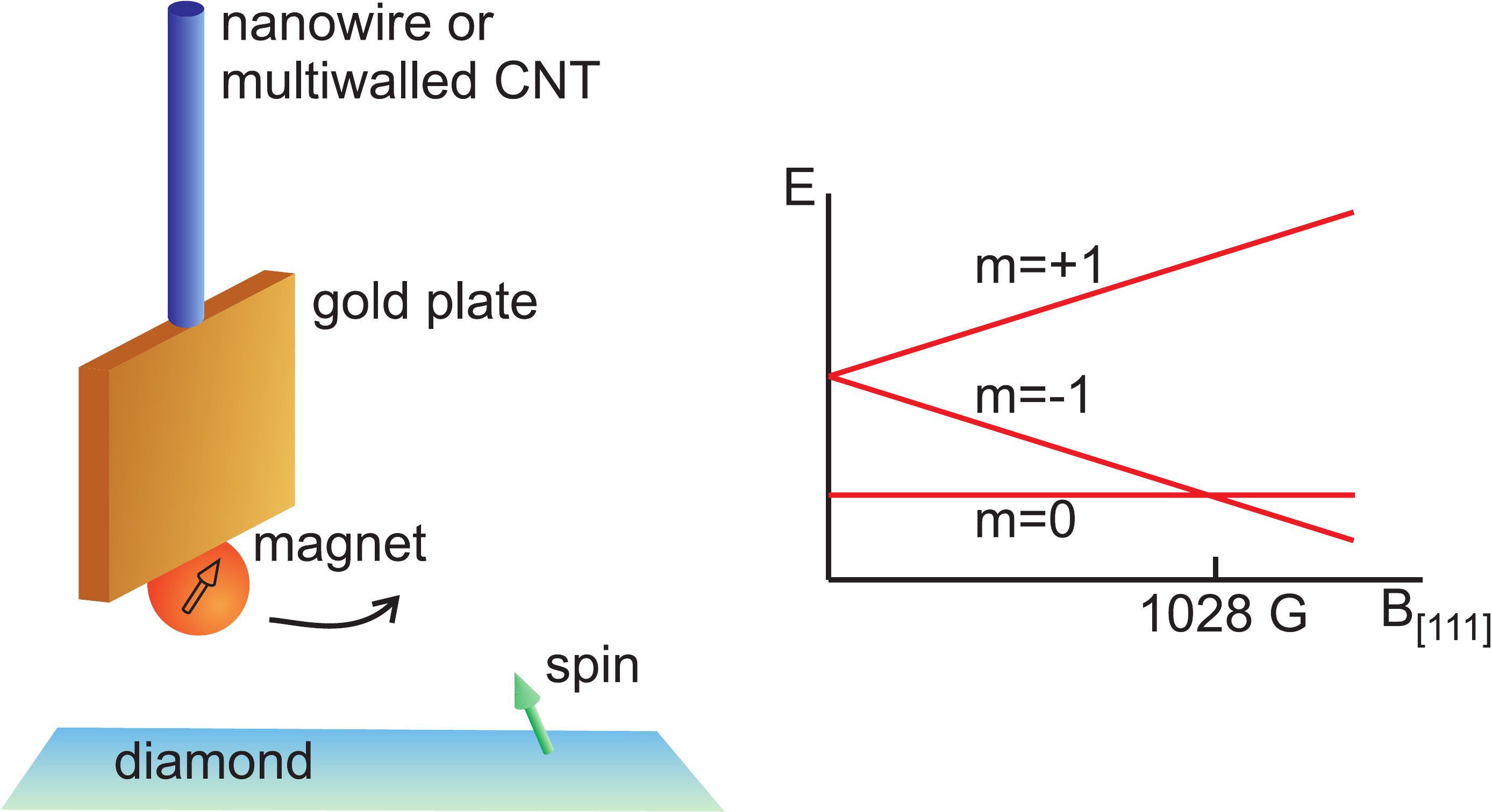}
      \end{center}
      \caption{The proposed experimental configuration. A thin gold plate is attached to a nanowire or multi-walled carbon nanotube to create a massive mechanical resonator. A small spherical magnet acts as a detector for the spin magnetic moment in a diamond NV centre. The graph on the right shows the energy levels of the spin as a function of magnetic field.}
      \label{resonator}
\end{figure}

Inspired by the experiment by Rugar et al. \cite{Rugar2}, in which the force exerted by an electron spin on a small magnet is detectable by measuring the deflection of a mechanical resonator, we have proposed that such an experimental configuration can be used to bring a significant mass in a superposition involving a large displacement \cite{vWezel:PhilMag}. The adaptation of such a setup shown in figure \ref{resonator} consists of a thin wire holding a plate of mass $m$ as well as a small spherical magnet with magnetization ${\bf M}$, in close vicinity to a single, isolated electron spin. A natural candidate for the electron spin is the well known Nitrogen-Vacancy (NV) color centre in diamond because already at room temperature it has shown a coherence time as large as $250$ microsec \cite{Takahashi}. In isotopically engineered diamond this increases to $1.8$ msec \cite{Wrachtrup}, while at cryogenic temperatures the $T_2$ time may be expected to become even larger. Rabl et al. have developed a purely quantum mechanical description of this mechanical resonator coupled to a single electron spin \cite{Rabl}. For independent detection of both the cantilever and the electron spin we envision that the cantilever motion can be observed by coupling it to a SQUID through a coil, while the electron spin can be characterized optically. When no current is injected into the SQUID and when no light is coupled to the NV centre, the purely quantum mechanical description of Rabl et al. should normally be applicable.

While the non-magnetic $|S^z \! = \! 0 \rangle$ ground state of the spin leaves the resonator untouched, the $|S^z \! = \! -1\rangle$ (and $|S^z \! = \! +1\rangle$) excited state will attract (and repel) the cantilever with a force $F=\mu G$ where G is the magnetic field gradient originating from the magnetic sphere at the position of the electron spin with magnetization $\mu$. Inducing transitions between and superpositions of the different spin states can readily be achieved by applying an appropriate combination of static $B_0$ and radio frequency $B_1$ magnetic fields. Starting from a superposed spin state, the magnetic coupling between the spin state and the motion of the resonator then eventually yields a superposition of two out of phase oscillation modes of the resonator.

For the experimental protocol described in detail in section \ref{section4}, and employing the realistic parameters discussed presently, the amplitude of the two anti-phase oscillations could be caused to approach the thickness of the plate. Importantly, these amplitudes could in principle be achieved in a time shorter than the expected dephasing time of the electron spin due to nuclear spins in the diamond lattice \cite{Wrachtrup} and shorter than the dephasing time of the cantilever due to coupling to the phonon bath \cite{Chen,Chen2}. Thus, a superposition of the plate over macroscopically distinct positions may be achieved within the limits set by decoherence.

\subsection{Order of Magnitude Estimates for Experimental Parameters}
The experimental parameters required to be able to unambiguously assign an observed decay time to any process other than decoherence lie only just beyond what is already being used in present day MRFM measurements. In the experiments of Rugar et al. the deflection of a resonator due to the interaction with a single electron spin was measured \cite{Rugar2}. The field gradient in these experiments was $\partial B / \partial x = 2\cdot 10^5$~T/m which, with a magnetic moment of the electron spin of $\mu_B=9.3 \cdot 10^{-24}$~J/T, leads to a force $F_{\text{spin}}=\mu_B \partial B / \partial x =1.8 \cdot 10^{-18}$~N. With the stiffness of the resonator $k=1.1 \cdot 10^{-4}$~N/m, this would imply a static deflection of only $d_0=F_{\text{spin}}/k=1.7 \cdot 10^{-14}$~m. 

To increase the deflection of the resonator, the electron spin was inverted twice during each resonator period, $T_{\text{res}}$. If the spin inversions remain coherent with the resonator motion long enough, the amplitude could in principle be increased to $d=Q F_{\text{spin}}/k$, where $Q$ is the Q-factor of the resonator. Since in practice the spin inversions remain coherent with the cantilever motion only for a time $\tau_m \ll Q T_{\text{res}}$, the maximum amplitude is limited to $d = (1-\exp(-\tau_m / Q T_{\text{res}}))\cdot Q F_{\text{spin}}/k$, where $\tau_m$ is the so-called rotating frame relaxation time. For the experiment by Rugar et al. $\tau_m = 760$~msec and the maximum amplitude would exceed $60$~pm \cite{Rugar}. 

For a spin in a quantum superposition of states, $\tau_m$ would have to be replaced by a dephasing time $T_2$, which has been measured to be $1.2$ msec in NV centers in diamond in which the $^{13}$C isotopic content was reduced \cite{Wrachtrup}. Although this is the longest decoherence time measured in a solid state system to date, the fact that $T_2$ is so much shorter than $\tau_m$ still limits the distance between the two centers of mass involved in the superposition. Fortunately, in the context of nuclear magnetic resonance force microscopy, the group of Rugar now employs field gradients of $4\cdot 10^6$~T/m \cite{Rugar}, which when applied to an electron spin experiment, would lead to a $20$ fold increase of the deflections. An alternative way of enhancing the deflection would be to employ softer springs. In our lab \cite{oosterkamp:private}, we have been able to fabricate nanowire resonators with spring constants down to $k=1 \cdot 10^{-6}$~N/m. Such a nanowire, ending in a thin gold foil measuring $5$~{$\mu$}m x $5$~{$\mu$}m x $20$~nm, would have a resonance frequency of $1.6$~kHz. 

Combining an increase of the NV-centre dephasing time to $T_2=50$~msec (which might be achievable by further eliminating nuclear spins in the diamond or by working at sufficiently low temperatures to freeze out the nuclear spins) with the highest field gradients in the Rugar lab and a soft $k=1 \cdot 10^{-6}$~N/m resonator would result in a deflection of $d = (T_2 / T_{\text{res}}) \cdot (\mu_B / k) \cdot \partial B / \partial x = 3$~nm. As explained in the next section, the typical energy scale involved in the process of spontaneous unitarity breaking may be expected to be $\Delta=G m^2 d^2 /L^3=4 \cdot 10^{-34}$~J (with $L$ the sample thickness and $G$ the gravitational constant), which yields a typical time scale $\hbar / \Delta = 25$~msec. 
Since this is of the same order of magnitude as the dephasing time of the NV-centre, it may be possible for the loss of unitarity to become apparent before the NV centre decoheres.

The dephasing of the cantilever itself, on the other hand, may be an issue. This time scale can be estimated by describing the environment as an infinite bath of harmonic oscillators and integrating out the environmental degrees of freedom \cite{ZurekPT,Luuk}. For the specific case where the dephasing is due to the phonons which are excited within the resonator, this problem has received some attention in the context of gravitational wave detectors and the timescale resulting after the phonon bath has been integrated out was shown to be $\tau_{\text{res}}=Q T_{\text{res}}/(2 \pi N_{\text{phon}})$, where $N_{\text{phon}}=k_B T_{\text{mode}} / h f_{\text{res}}$ is the number of phonons of the resonator mode, which is determined solely by the resonator's thermodynamic mode temperature $T_{\text{mode}}$, and frequency $f_{\text{res}}$. In order for this dephasing time to exceed $\hbar / \Delta$, the temperature of a resonator with resonance frequency $f_{\text{res}}$ and $Q=10^5$ would have to be smaller than $100$~{$\mu$}K, which would require the experiment to be done in a nuclear demagnetization refrigerator. However, although before we considered the damping $\gamma = k/(2\pi f Q)=1 \cdot 10^{-15}$~Ns/m with $Q=10^5$, it is possible that future resonators will have even higher $Q$ since much lower damping values have been achieved in the carbon nanotube resonators fabricated by Huttel et al. \cite{Huttel}, who estimate $\gamma=2\pi f_{\text{res}} m/Q=2\cdot 10^{-17}$. If such a low damping can be achieved for the resonator proposed here, the mode temperature may be around $1$~mK without dominating the dephasing time of the NV-centre.

It has been argued before that the thermal contribution to the dephasing time considered above should give a correct order of magnitude estimate even in the presence of a driving force \cite{Bouwmeester}. However, if we consider a dephasing mechanism whose strength depends explicitly on the amplitude of the oscillation as well as its mode temperature \cite{Unruh}, we should replace $N_{\text{phon}}$ in the expression for $\tau_{\text{res}}$ by $\frac{1}{2} k d^2 / h f_{\text{res}}$. With the lowest damping considered ($\gamma=2\cdot 10^{-17}$) at a temperature of $100$~$\mu$K, this results in a dephasing time of the order of $1$~msec. In that case, one would need to resort to using an even thinner gold plate to see the effect of the spontaneously broken unitarity. A gold mass of $10$~nm thickness (f.e. deposited on a single sheet of graphene), in the presence of such low damping, gives rise to the timescale $\hbar / \Delta = 20$~msec, which might just be detectable within the time limit set by decoherence.

Although the phonon contribution considered here is expected to dominate the dephasing of the resonator, other mechanisms are also present. Defects in the cantilever for example may act as effective two-level systems at low temperatures, and the clamping of the cantilever may lead to additional phonon radiation, both contributing to the dephasing of the resonator \cite{Remus}. The former issue can be addressed by the application of small additional magnetic fields or by reducing the number of defects in the carbon nanotube resonator, while a scheme to circumvent the problem of clamping loss using optical trapping of a resonator has recently been proposed in a different context \cite{Kheifets}.

\section{Simulating the Time Evolution}
\label{section3}
In this section, we will focus on spontaneously broken unitarity as one particular scenario for the transition from quantum dynamics to classical physics, and show explicitly how its predictions may be expected to influence the dynamics of the coupled resonator-spin system. How the proposed experimental setup may be used to also distinguish these particular predictions from those of other scenarios will be discussed in the next section. A detailed discussion of the theory of spontaneously broken unitarity can be found in ref. \cite{vWezel:Review}, and we will only repeat its main results here. 

The basic observation underlying the idea of spontaneous unitarity breaking is the fact that the dynamics generated by the usual time dependent Schr\"odinger equation is unstable, in the sense that even an infinitesimally weak perturbation may qualitatively change its behavior in the thermodynamic limit. This situation is analogous to that of other spontaneously broken symmetries: the Hamiltonian for a crystal is invariant under translations, but the sensitivity to even infinitesimal perturbations allows a macroscopic crystal to nonetheless localize in only a single position \cite{PWA}. In the same way the unitarity of quantum mechanical time evolution is sensitive to even infinitesimally small non-unitary perturbations, which allows sufficiently large objects to undergo classical dynamics \cite{vWezel:Review}. An important caveat in this argument is that there has to exist a fundamental non-unitary interaction somewhere in nature \cite{vWezel:PRB}. It has been pointed by several authors that gravity may fill this role, and that in fact the energy scale on which the effects of gravity as a non-unitary influence become important lies in the regime in which we expect the quantum to classical crossover to take place \cite{Diosi,Penrose,vWezel:PhilMag}.

Because of the relatively small mass and low density of the resonator in the proposed experiment, its dynamics will  be very close to purely quantum mechanical, and we expect to be able to include any non-unitary effects due to gravity as minor perturbations to Schr\"odinger's equation. Following the procedure outlined in the appendix, we thus write the dynamics of the resonator in the presence of a non-unitary gravitational perturbation as:
\begin{align}
\frac{d}{d t} \psi = - \frac{i}{\hbar} \left( \hat{H} - i  G \frac{m^2}{2 L^3} \left[\hat{x}-\xi\right]^2 \right) \psi,
\label{MSE}
\end{align}
where $\hat{H}$ is the usual quantum mechanical Hamiltonian, $G$ is the gravitational constant, and $m$ and $L$ are the mass and width of the massive plate. The operator $\hat{x}$ is the position operator for the centre of mass of the plate (measured with respect to the centre of mass of the initial wavefunction). The time-dependent, randomly fluctuating variable $\xi$ has been introduced as a correction to $\hat{x}$, because the theory of gravity (i.e. general relativity) insists that any quantity which can be used as a measure of distance between locations in different components of a spacetime superposition must be ill-defined. 

Notice that both general relativity and quantum mechanics are in fact fully deterministic theories. There is thus no reason to believe that any part of the interplay between quantum mechanics and gravity should be anything but deterministic. The introduction of a random variable in equation \eqref{MSE} should be seen only as a poor man's approach towards simulating an essentially ill-defined quantity: close to the point where unitary quantum mechanics is still a good description of nature we may get away with using the concept of superpositions --even though they really are ill-defined notions in general relativity-- if we include also an effectively random correction to the differences in distance between superposed spacetimes. Although superpositions and random variables may not actually feature in the exact reality of quantum gravity, we assume that if we insist on the possibility of {\it effectively} describing the state of the system as a superposition in some limit, then we also need to take into account an {\it effectively} random correction to the notion of position. We are thus led to the effective, phenomenological description of the  first order correction to the unitary Schr\"odinger equation given by equation \eqref{MSE}. In the appendix we give a more detailed discussion of both the steps leading to equation \eqref{MSE} and the implications of the imaginary term on the conservation of energy and on the conservation of the norm of the wavefunction.

In contrast to the Schr\"{o}dinger-Newton equation, which has been formulated in a similar context by other authors \cite{SchN,Jeroen}, equation \eqref{MSE} is a purely linear equation, which arises naturally from the extension of the well studied equilibrium mechanism of spontaneous symmetry breaking to the dynamical realm, and obeys all the requirements of a theory of spontaneous symmetry breaking: for unitarity to be spontaneously broken, one necessarily needs to invoke a singular thermodynamic limit, a macroscopic order paramter, and a symmetry breaking field \cite{vWezel:Review}. It is because of this connection to spontaneous symmetry breaking, that the timescale $G m^2 \langle x^2 \rangle /2 L^3$, which has been suggested before to set the appropriate time scale for the influence of gravity on quantum mechanical time evolution \cite{Diosi,Penrose,vWezel:PhilMag}, may be embedded into the explicit expression of the wavefunction dynamics provided by equation \eqref{MSE}.
This final form of the modified Schr\"odinger equation can be straightforwardly integrated using standard numerical methods to yield a phenomenological prediction for the expected time evolution of our resonator experiment.

\subsection{Two Time Scales}
To be explicit, we first express the state of the system depicted in figure \ref{resonator} by the quantum numbers for the position of the cantilever's centre of mass $x$ and the orientation of the spin's magnetic moment $\sigma$. We are then interested in the dynamics of the system starting from the initial state $\varphi(x,t \! = \! 0)=\alpha \chi_{\uparrow} \psi^0(x+d) + \beta \chi_{\downarrow} \psi^0(x-d)$, where $\psi^0(x)$ is the groundstate wavefunction of the harmonic oscillator of mass $m$ centered at $x=0$, and $\chi_{\sigma}$ indicates the spin state with $S^z=\sigma$. The entangled initial state is prepared by performing an MRFM measurement on a suitably initialized superposition of the spin state, as discussed in the next section.

 For the purpose of clarity we ignore the imposed oscillatory motion of the cantilever in this simulation and assume instead that it is fixed at its maximum displacement $\pm d$. It is straightforward to also include these oscillations but they shroud the role of the unusual dynamics induced by the phenomenological gravitational term and do not change our conclusions. Further ignoring for the moment the phonons and other internal degrees of freedom of the oscillator, we are left with only the kinetic energy of the oscillator as a whole to set the Hamiltonian in equation \eqref{MSE}, so that we have $\hat{H} = \hat{p}^2 / ( 2 m )$. The spin in the diamond NV centre is assumed to be free for the duration of the experiment.
\begin{figure}
      \begin{center}
      \includegraphics[width=0.75\columnwidth]{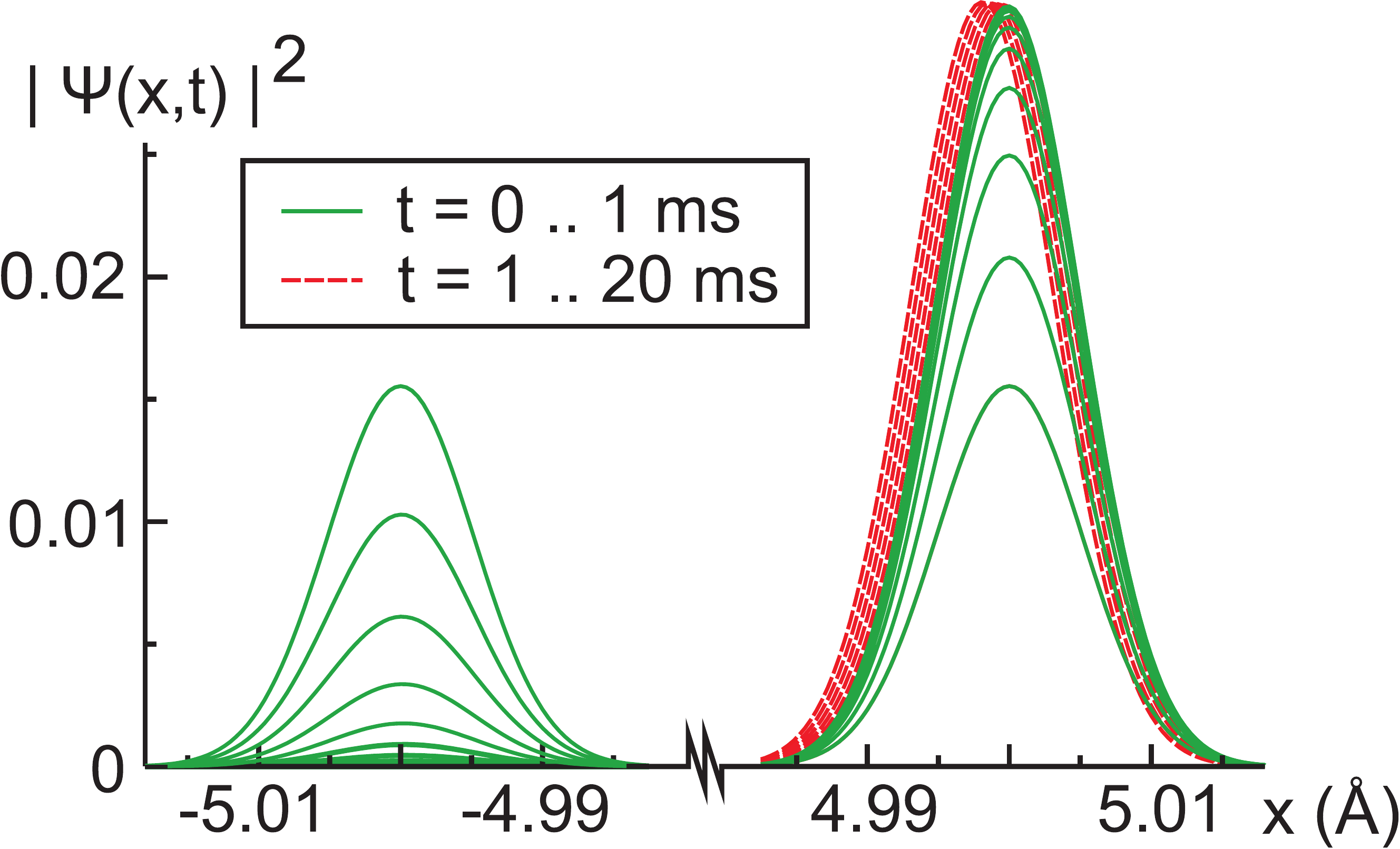}
      \end{center}
      \caption{The two reduction processes induced by the modified Schr\"odinger equation. Starting at $t=0$ with the superposed wavefunction $\sqrt{1/2} \ ( \chi_{\uparrow} \psi^0(x+d) + \chi_{\downarrow} \psi^0(x-d) )$ and $\xi > 0$, the time evolution induced by equation \eqref{MSE} with a static $\xi$ initially causes the fast reduction of the wavefunction to just the component $\chi_{\downarrow} \psi^0(x-d)$ (solid curves), and then the slow reduction of that component towards a wavefunction centered at $x=\xi$ (dashed curves). For clarity, all curves have been renormalized before plotting them. The parameters used in this simulation are based on the optimistic order of magnitude estimates discussed in section \ref{section2}. They are $m=1 \cdot 10^{-14}$~kg, $L=1$~nm and $d=1$~nm, resulting in an initial reduction time of the order of $t=1$~ms.}
      \label{plotA}
\end{figure}

The time-dependent wavefunction of the resonator found by solving the differential equation \eqref{MSE}, now shows two distinct processes which happen simultaneously. These can be most easily understood for the special case of constant $\xi(t)$. In that case the gravitational term of equation \eqref{MSE} exponentially suppresses the weights of both components of the wavefunction, but one component is suppressed more strongly than the other, depending on whether $x-\xi$ is positive or negative. One component thus quickly dominates the overall wavefunction, and the superposition is reduced to just the single product state $\chi_{\uparrow} \psi^0(x+d)$ or $\chi_{\downarrow} \psi^0(x-d)$, as is shown in figure \ref{plotA}. That the reduction of the superposed wavefunction results in just one component in the position basis is a direct consequence of gravity providing the unitarity breaking field in equation \eqref{MSE}. Because of the special role played by the position variable in the theory of gravity, spatial superpositions become unstable and the pointer basis which defines the possible outcomes of the reduction process consists of spatially localized states. Notice that the spatial reduction of the cantilever wavefunction automatically implies that also the entangled spin state will end up in only one of its components, even though it is not itself subject to any gravitational effects.

The localization dynamics is taken even one step further in the second process induced by the modified Schr\"odinger equation. Because the single component $\psi^0(x \pm d)$ is itself a wavefunction spread out in space, it too is subject to the uneven suppression by the gravitational term. As long as the spread of the wavefunction (i.e. its deBroglie wavelength) is short compared to the distance $\xi \pm d$ however, the difference in amplification rates of its components is very small, and the secondary reduction of the wavefunction into a state centered at $x=\xi$ will be very slow (see figure \ref{plotA}).
\begin{figure}
      \begin{center}
      \includegraphics[width=0.75\columnwidth]{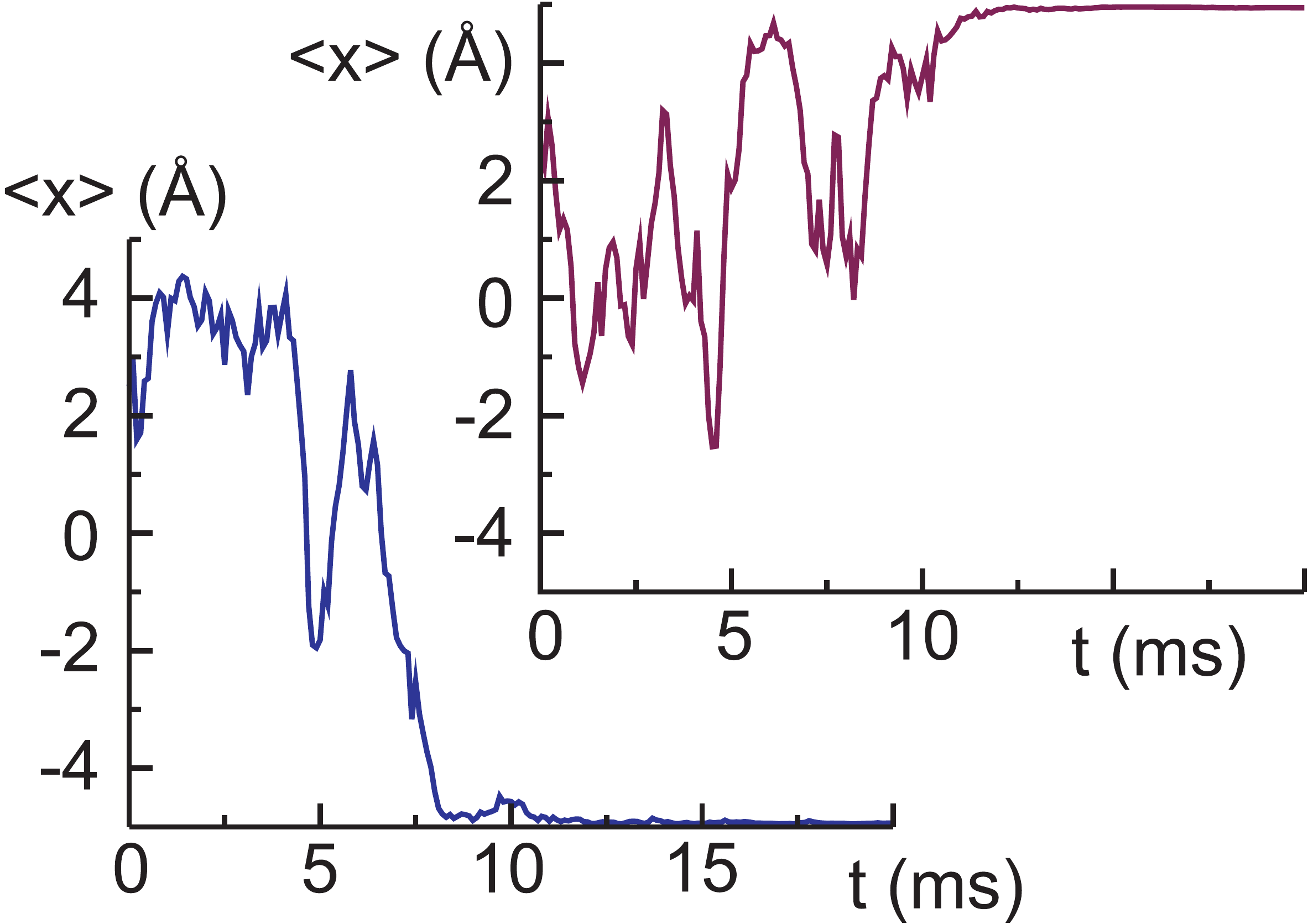}
      \end{center}
      \caption{Two typical realizations of the dynamics of equation \eqref{MSE} in the presence of a randomly fluctuating $\xi(t)$, starting from the initial wavefunction $\sqrt{1/4} \ \chi_{\uparrow} \psi^0(x+d) + \sqrt{3/4} \ \chi_{\downarrow} \psi^0(x-d)$. The fate of the wavefunction is tracked by the expectation value of the position operator, $\langle \hat{x} \rangle$. The initial wavefunction is reduced to just one of its components in a time proportional to the inverse of the strength of the gravitational term in equation \eqref{MSE}. The parameters used in these simulations are the same as in figure \ref{plotA}.}
      \label{plotB}
\end{figure}

\subsection{Born's Rule}
In the presence of a fluctuating $\xi(t)$, the same two processes occur. This time however $x-\xi$ may alternate in time between being positive or negative. It will thus alternately amplify the relative weights of the two different components of the initial superposition, and the question of which component wins out in the end becomes a realization of the ``gambler's ruin'' game: each component will randomly increase and decrease in relative weight until one of them completely dominates. Although the fluctuations of the stochastic variable $\xi(t)$ slow down the rise to dominance of either of the components, a final resolution can still be seen to be reached within a typical timescale proportional to the inverse of the gravitational energy $\Delta \equiv G m^2 d^2 /L^3$ defined in equation \eqref{MSE} \cite{vWezel:Review}, as expected on general dimensional grounds \cite{vWezel:PhilMag,Penrose,Diosi}. This energy scale sets the minimum time scale required by the resonator dynamics to unambiguously display non-unitary effects.

\begin{figure}
      \begin{center}
      \includegraphics[width=0.65\columnwidth]{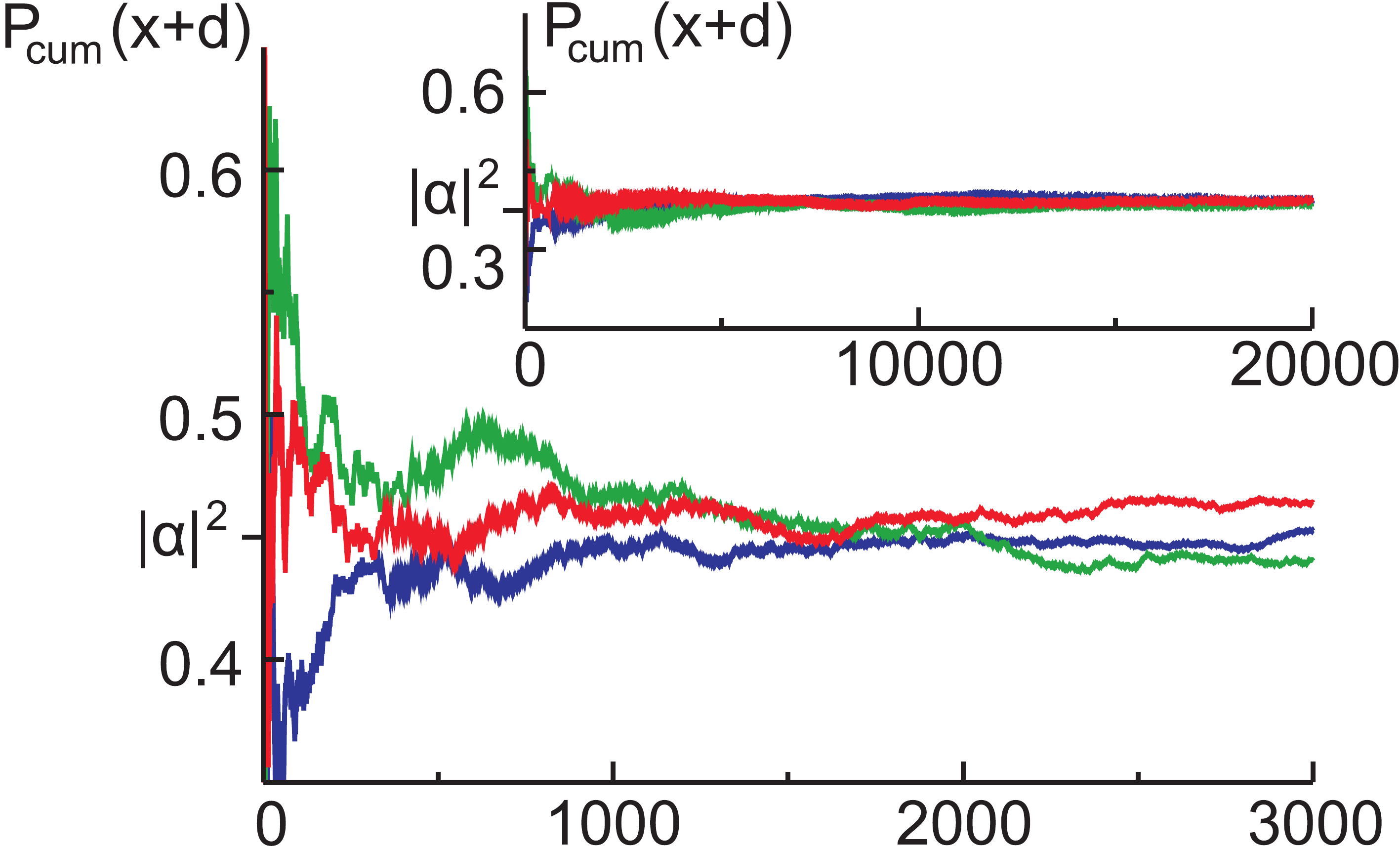}
      \end{center}
      \caption{The cumulative proportion of wavefunctions being reduced to $\chi_{\uparrow} \psi^0(x+d)$, starting from $\alpha \ \chi_{\uparrow} \psi^0(x+d) + \beta \ \chi_{\downarrow} \psi^0(x-d)$. The initial dynamics shown in figure \ref{plotB} was repeated  $20000$ times for each of three different sets of the parameters $m$ and $d$. For each set the cumulative proportion of outcomes resulting in $\chi_{\uparrow} \psi^0(x+d)$ is seen to be $| \alpha |^2$, in accordance with Born's rule. Here we used $\alpha=\sqrt{9/20}$.}
      \label{plotC}
\end{figure}

An example of a typical solution of the modified Schr\"odinger equation with a fluctuating stochastic variable is shown in figure \ref{plotB}. 
Because $\xi(t)$ is a randomly oscillating variable, it is impossible to predict which of the two components will survive the imposed dynamics. If one of the components has a larger weight in the initial wavefunction however, it is more likely to survive. In fact it is possible to mathematically prove that after an infinitely long time the only possible average outcome of this process is the emergence of Born's Rule \cite{vWezel:PRB,Zurek}. That is, if we repeat the simulation many times always starting from the same initial state $\alpha \chi_{\uparrow} \psi^0(x+d) + \beta \chi_{\downarrow} \psi^0(x-d)$, but randomly generating a new $\xi(t)$ for each run, the average fraction of evolutions resulting in the final state $\chi_{\uparrow} \psi^0(x+d)$ converges to $|\alpha|^2$, as in figure \ref{plotC}.

\section{Protocol for the Experiment}
\label{section4}
Returning now to the experimental setup described in section \ref{section2}, we need to find a way of distinguishing the dynamics of spontaneous unitarity breaking as described by equation \eqref{MSE} from both the predictions of alternative models of the quantum to classical crossover, and from the ubiquitous decoherence effects. Especially the latter is not an easy task, because in the proposed setup, it is not possible to directly observe the time evolution of the state of the cantilever as displayed in figure \ref{plotB}, and we have to rely on ensemble measurements instead.

\subsection{Distinguishing Decoherence from Decay} 
The central observation in the theory of decoherence is that the many unobservable microscopic degrees of freedom which are necessarily present in any experiment, have the effect, if averaged over an ensemble of measurements, to reduce any initially pure density matrix to a classical mixture of pointer states \cite{decoherence}. Although decoherence and the non-unitary dynamics of equation \eqref{MSE} have completely different origins and differ substantially in their descriptions of a single measurement, the ensemble averages look very similar. In terms of the reduced density matrices (i.e. density matrices averaged over all unobservable environmental degrees of freedom), both give rise to an effective, ensemble-averaged evolution of the type
\begin{align}
\rho(t) = \left( \begin{array}{cc} |\alpha|^2 & \alpha \beta^* e^{- t / \tau} \\ \alpha^* \beta e^{-t / \tau}  & |\beta|^2 \end{array} \right).
\label{rho}
\end{align}
Here $\tau$ is either the decoherence time $\tau_{\text{decoh}}$ or the average decay time $\tau_{\text{decay}}$. To distinguish the  influence of the non-unitary term from the more usual effects coming from various sources of decoherence, we propose an experimental protocol consisting of two different stages. 

\subsubsection{Stage One}
In the first instance, the spin and the resonator should be entangled in such a way as to give rise to a spatial superposition of resonator states. This can be achieved in a standard Magnetic Resonant Force Microscopy experiment in which the spin is reversed at twice the resonator frequency to drive the resonator, but starting from a superposed spin state. One component of the spin wavefunction will then enhance the resonator motion more strongly than the other, and as long as coherence can be maintained, the result will be an entangled state in which the resonator motion has a larger amplitude in one component than in the other:
\begin{align}
\varphi(t=0) &= \alpha \chi_{\uparrow} \psi^L(x) + \beta \chi_{\downarrow} \psi^S(x) \notag \\
\Leftrightarrow \rho(t=0) &= \left( \begin{array}{cc} |\alpha|^2 & \alpha \beta^* \\ \alpha^* \beta  & |\beta|^2 \end{array} \right),
\label{t0}
\end{align}
where $\psi^{L}(x)$ and $\psi^S(x)$ indicate the wavefunctions of the oscillator with enhanced or suppressed amplitude respectively, and the spin's $S^z$ component in $\chi_{\sigma}$ is measured in the corotating frame. This state with spatially separated components for the massive resonator will be sensitive to gravity-induced non-unitary effects as well as to decoherence, so that the off-diagonal element of its reduced density matrix after a given time interval $t_1$ is given by:
\begin{align}
\rho_{12}(t_1) = \alpha \beta^* e^{- t_1 / \tau_{\text{decoh}} - t_1 / \tau_{\text{decay}}}.
\label{t1}
\end{align}

The effective dephasing time (which combines the gravitational decay time and the effects of decoherence) can be found after many repetitions of the experiment in which the final state is read out by either coupling the nanomagnet attached to the oscillator to a SQUID through a coil, or alternatively by measuring the fluorescence coming from the spin at the NV centre after the application of a microwave $\pi/2$ pulse (which rotates the spin orientation into the $xy$ plane). The effective dephasing time found in this way, in general depends on both the mass and the shape of the resonator. The characteristics of this dependence can be used as a first means of distinguishing gravity-induced non-unitary dynamics from any given, known source of decoherence. For any given source of decoherence it is straightforward to work out the functional dependence of $\tau_{\text{decoh}}$ on the mass of the resonator, its linear size, its geometric shape, etc. In general, one or more of these dependencies will differ from the ones implied by the non-unitary form of equation \eqref{MSE}. For example, because they depend only on internal degrees of freedom of the oscillator, most sources of decoherence are independent of the angle between the direction of oscillation and the normal to the surface of the gold plate. By measuring the functional dependence of the observed  $\tau$ on such parameters, we can thus rule out most known sources of decoherence.
To make sure that no {\em unknown} sources of decoherence are at play either, one can then add a second stage to the experiment.

\subsubsection{Stage Two}
Any source of decoherence, no matter what its precise physical origin is, will arise from the entanglement of the cantilever motion with some other, unobserved degrees of freedom (called the bath). The gravitational decay on the other hand, only requires the presence of a massive superposition, and takes place even in complete isolation. To differentiate between the two mechanisms, one thus needs to ascertain that the cantilever is not entangled with any bath degrees of freedom, known or unknown. This is done in the next state, where we again start by creating the entangled superposition of equation \eqref{t0}, but rather than reading out the wavefunction of the resonator after a given time interval, we imprint its state back onto the diamond spin. 

To do this, one could use a spin echo setup, and flip the spin at a given instant in time (using an extra microwave $\pi$ pulse, which inverts the orientation of the spin), so that from then on the magnetic force between spin and resonator will tend to damp its motion if it enhanced it before the spin flip, and {\it vice versa}. Thus the effect of the different spin orientations on the oscillator position will be effectively undone. Once the resonator has traced its way back to its original wavefunction, the position of the resonator and the spin direction are no longer entangled and the magnetic driving field can be switched off. In the absence of any decoherence and decay processes, the evolution of the wavefunction, including the creation of the entangled state, would be simply:
\begin{align}
\varphi(t=0) &= \left[ \alpha \chi_{\uparrow} + \beta \chi_{\downarrow} \right] \psi^0(x) \notag \\
\varphi(0<t<t_1) &= \alpha \chi_{\uparrow} \psi^L(x,t) + \beta \chi_{\downarrow} \psi^S(x,t) \notag \\
\varphi(t_1<t<t_2) &= \alpha \chi_{\downarrow} \psi^L(x,t) + \beta \chi_{\uparrow} \psi^S(x,t) \notag \\
\varphi(t=t_2) &= \left[ \alpha \chi_{\downarrow} + \beta \chi_{\uparrow} \right] \psi^0(x).
\end{align}
Here the spin echo pulse was applied at $t=t_1$, and the time dependence of $\psi^{L,S}(x,t)$ signifies a growing (or shrinking) oscillation amplitude as a function of time for $t<t_1$, but the opposite for $t>t_1$, due to the opposing influence of the flipped spin. If a non-unitary term is present, the components $\alpha$ and $\beta$ become time dependent, but the oscillator state will still be brought back to $\psi^0(x)$ at $t=t_2$. Likewise, decoherence may result from the entanglement of these states with bath degrees of freedom, but these do not alter the final oscillator state.

After decoupling the spin state and the oscillator state at $t_2$, there will not be any additional gravitational decay because there is no longer any spatially superposed mass. Any decoherence caused by the entanglement of the diamond spin with unknown internal (bath) degrees of freedom of the resonator however, will continue unabated even if the resonator remains at rest. After all, due to the rigidity of the resonator (and the associated Goldstone theorem), its internal degrees of freedom are decoupled from its collective coordinates. Even though the spin flip at $t=t_1$ effectively reverses the time evolution of the collective, centre-of-mass motion of the resonator, the dynamics of the internal, bath degrees of freedom is not reversed, and they remain in an entangled state even after $t=t_2$. The full wavefunction evolution in the presence of both decay and bath degrees of freedom can thus schematically be written as:
\begin{align}
\varphi(t=0) = & \left[ \alpha(0) \chi_{\uparrow} + \beta(0) \chi_{\downarrow} \right] \psi^0(x) \phi^{0}(0) \notag \\
\varphi(0<t<t_1) = & \alpha(t) \chi_{\uparrow} \psi^L(x,t) \phi^{L}(t) + \beta(t) \chi_{\downarrow} \psi^S(x,t) \phi^{S}(t)  \notag \\
\varphi(t_1<t<t_2) = & \alpha(t) \chi_{\downarrow} \psi^L(x,t) \phi^{L}(t) + \beta(t) \chi_{\uparrow} \psi^S(x,t) \phi^{S}(t) \notag \\
\varphi(t > t_2) = & \left[ \alpha(t_2) \chi_{\downarrow} \phi^{L}(t) + \beta(t_2) \chi_{\uparrow} \phi^{S}(t) \right] \psi^0(x).
\end{align}
Here $\phi^{L,S}(t)$ represent the evolution of the bath degrees of freedom, as influenced by the different amplitude states of the oscillator. Their time evolution continues beyond $t=t_2$, while the components $\alpha$ and $\beta$ become time-independent after the centre of mass state has returned to $\psi^0(x)$. After tracing out the bath degrees of freedom and simultaneously averaging over many realizations of the stochastic dynamics imposed by the non-unitary time evolution, the off-diagonal element of the reduced density matrix for the final state yields:
\begin{align}
\rho_{12}(t>t_2) = \alpha(0) \beta(0)^* e^{-t_2 \left( 1 / \tau_{\text{decoh}} + 1/ \tau_{\text{decay}} \right)} e^{- \left(t-t_2\right) / \tau_{\text{decoh}}}.
\label{t2}
\end{align}

At the end of the second experimental stage then, the coherence of the diamond spin can again be monitored via optical readings in many repetitions of the experiment. The initial suppression of the off-diagonal matrix elements by the factor $e^{- t_2 / \tau_{\text{decoh}} - t_2 / \tau_{\text{decay}} }$ is known from the first stage of the experiment. The remaining suppression of the off-diagonal matrix elements by the factor $e^{- (t-t_2) / \tau_{\text{decoh}}}$ can be attributed purely to sources of decoherence. It will most likely be dominated by the well-characterized decoherence of an isolated diamond spin due to sources within the diamond environment itself. For the experimental parameters proposed here, this timescale should be in the millisecond range, and can be easily recognized. Any remaining sources of decoherence which had not been identified and eliminated in the first experimental stage can now be recognized as an apparent contribution to $e^{- (t-t_2) / \tau_{\text{decoh}}}$, in excess of that expected from the known, pristine diamond environment. Once these remaining unidentified bath degrees of freedom have been eliminated (f.e. through additional cooling), the dephasing measured in the first stage of the experiment can be conclusively attributed to a decay process instead of decoherence.

\subsection{Distinguishing Different Decay Models}
After ruling out decoherence as the source for the suppression of off-diagonal matrix elements, we still need to ascertain whether the observed effective decay time $\tau$ is due to the process of spontaneous unitarity breaking, which we focussed on in this paper, or to some other non-unitary mechanism. The most well-known alternatives of this kind fall in the GRW and CSL classes \cite{GRW,CSL0,CSL}. Other proposed scenarios for the quantum to classical crossover, such as hidden variable and many worlds theories are fully unitary \cite{hiddenvars,manyworlds}, and the observation of any non-zero decay time in the proposed experiment would thus suffice to rule out their relevance in the description of the resonator dynamics.

The main assumption in both the CSL and the GRW theories is the existence of a specific process beyond the realm of applicability of Schr\"odinger's equation. In the GRW theory this process is assumed to be the spontaneous and instantaneous spatial localization of an elementary particle at particular intervals in time \cite{GRW}. If the frequency with which these localization events occur is low enough, it will take an immeasurably long time (on average) for any individual particle to undergo such an event. Within an extended object consisting of a macroscopic number of particles on the other hand, it will not take long before at least one of the particles is localized. Assuming that the extended object possesses some degree of rigidity, the localization of a single particle within it will suffice to give the entire wavefunction a definite position in space. Macroscopic objects are thus rapidly reduced to position eigenstates, while microscopic particles are free to spread out in space. Born's rule can be built into this construction by making additional assumption about the localization process \cite{GRW}. 

Because the localization process proposed by GRW acts on individual microscopic particles, and does not depend on the overall state of the system, its predicted decay time scales only with the number of particles involved in the massive superposition. A dependence of the decay time $\tau_{\text{decay}}$ in equation \eqref{t2} on the involved mass or shape of the resonator will thus clearly indicate a departure from the predictions of the GRW theory, and require additional non-unitary effects.

The CSL (or Continuous Spontaneous Localization) models can be seen as extensions of the original GRW model in which the addition to quantum mechanics is no longer a set of instantaneous localization events, but rather a continuous process which constantly acts to gradually localize the individual particles. In most modern CSL models, a smeared mass density is taken to be the variable which specifies both the rate and the final states of the reduction process \cite{CSL0,CSL,CSL2}. The obtained average rate of localization ensures that microscopic particles take an immeasurably long time to localize, while macroscopic objects again localize almost instantaneously due to their rigidity. Born's rule can emerge spontaneously if the localization process is stochastic, which is modeled by the inclusion of white noise in the dynamical equations \cite{CSL0}.

The final form of the reduction process described by the CSL model resembles the dynamics of spontaneous unitarity breaking in equation \eqref{MSE} in various ways. Both predict a reduction of the superposed initial state within a timescale proportional to the square of the total mass of the involved object; both involve a random variable which ensures that Born's rule can be obeyed; and both mechanisms select the position basis as the pointer basis to which macroscopic objects must be reduced. 

However, there are also clear differences between the two approaches. For example, as discussed in the review articles of Bassi and Ghirardi \cite{CSL0}, and of Pearle \cite{CSL2}, the effective reduction rate in the CSL model is set directly by the spread of the mass density distributions in different components of the initial wavefunction, independent of the relative locations of their centres of mass (if the distributions are initially well separated). Thus, in contrast to equation \eqref{MSE} the reduction rate of a macroscopic superposition is independent of the relative distance between the massive bodies in its components, as long as they are initially non-overlapping. Such a difference would be obvious in the experiment, if we could construct a superposition of the massive plate over distances greater than its thickness. If on the other hand the two components of the resonator wavefunction do have a substantial spatial overlap throughout the experiment, we need to rely on the dependence of the decay time on the precise geometry of the massive plate to differentiate between the predictions of the CSL model and spontaneous unitarity breaking. In the CSL model, the effective decay time can be shown to be proportional to $d / L^2$, where $d$ is the thickness of the non-overlapping part of the involved mass distributions \cite{CSL0}. From equation \eqref{MSE}, it can be seen that spontaneous unitarity breaking predicts the rate to be proportional to $d^2 / L^3$ instead \cite{vWezel:PhilMag}.

After ruling out decoherence as the cause for the decay of the off diagonal matrix element in the two-stage protocol, we can thus use the dependence of the observed decay time on the mass and geometry of the proposed setup to conclusively distinguish the effects of all currently proposed descriptions of the quantum to classical crossover.

\section{Conclusions}
We have proposed a nanoscale experiment, realizable with present-day technology, which can be used to produce an entangled state of a Nitrogen-Vacancy spin in diamond and a mechanical resonator.  Because the entangled state involves the superposition of a massive cantilever over a sizable distance, we are forced to consider the effects that a quantum to classical crossover might have on the dynamics of this state. The proposed experiment is focussed on testing one particular scenario for this crossover, which describes the disappearance of unitary quantum dynamics in the thermodynamic limit as a form of spontaneous symmetry breaking, akin to the spontaneous symmetry breaking observed in macroscopic crystals, magnets, etc.
By constructing an effective, phenomenological description of the proposed experiment, it is shown that the predicted effects of spontaneous unitarity breaking on the time evolution of the resonator state can be experimentally tested. A specific experimental protocol is proposed in order to distinguish the effects of spontaneous unitarity breaking from those of both decoherence and other proposed models for the quantum to classical crossover. 

The phenomenological description of the resonator dynamics predicts the reduction of a superposed initial state to just one of its components to take place on a timescale which lies just within experimental reach. The reduction of the cantilever state also implies the reduction of the entangled spin state, even though the spin itself is not directly under the influence of the broken unitarity. The emergence of both a pointer basis and Born's rule for the reduction process can be seen to be a direct consequence of the influence of gravity on quantum mechanics.

\section{Appendix A: The Norm of the Wavefunction}
To see how the usual probabilistic predictions of quantum mechanics may emerge from the modified time evolution of equation \eqref{MSE}, even though it does not conserve the normalization of the wavefunction \cite{vWezel:Review}, consider a particular run of the proposed experiment, with the spin of the diamond NV centre initially in a superposition of $S^z$ states, while the resonator is in a single position state: 
\begin{align}
\varphi(t < 0) = \left[ \alpha \chi_{\uparrow} + \beta \chi_{\downarrow} \right] \psi^0(x).
\end{align}
Here, without loss of generality, we assume $|\alpha|^2+|\beta|^2=1$. This state is a stable state, in the sense that the diamond NV centre is not affected by the non-unitary term of the modified Schr\"odinger equation since it does not involve a massive superposition, while the oscillator is in a state with a single, well-defined position, and is thus subject only to the extremely slow dynamics indicated by the dashed curves in figure \ref{plotA}. For the estimated experimental parameters discussed in section II, the slow motion of the oscillator happens on an unmeasurably long timescale. If we were to consider even heavier objects (like for example the pointer on a regular measurement machine), this time scale becomes even longer. For all practical purposes the product state $\varphi(t<0)$ is thus a static configuration, with a conserved norm.

In the initial phase of the experiment, the spin state and the oscillator wavefunction become entangled by the coupling of the spin orientation to the deflection of the cantilever in the usual MFRM setup. If this can be done within a time that is short compared to both the intrinsic coherence time of the spin and the typical decay time imposed by the non-unitary field, the resulting state will be given by:
\begin{align}
\varphi(t=0) = \alpha \chi_{\uparrow} \psi^0(x+d) + \beta \chi_{\downarrow} \psi^0(x-d).
\end{align}
This state is an entangled state, which involves the superposition of a massive object over different locations. It is therefore subject also to the faster non-unitary decay dynamics (indicated by solid curves in figure \ref{plotA}). For the experimental parameters proposed in section II, the fast non-unitary dynamics takes milliseconds to complete, and is within the measurable regime. For heavier objects, this time scale quickly becomes unmeasurably short. If we were to consider a truly macroscopic superposition (involving for example the pointer on a regular measurement device), the decay would, for all practical purposes, be instantaneous. 

During the decay dynamics, the norm of the wavefunction is not conserved. The result, if the decay happens on a time scale short compared to the involved intrinsic decoherence times, will be either one of two possible outcomes:
\begin{align}
\varphi(t \gg 0) = \left\{ \begin{array}{lr} & N \chi_{\uparrow} \psi^0(x+d)  \\ 
\text{\emph{or}} \ & \\
& N \chi_{\downarrow} \psi^0(x-d), \end{array} \right.
\label{final}
\end{align}
where the norm $N$ depends on the precise dynamics, and is certainly not equal to one. Despite the lack of a normalized wavefunction, there can be no doubt about the physical interpretation of the final states in equation \eqref{final}. Each of the possible outcomes is a single product state which combines an eigenfunction of the $z$-projection operator of the spin in the diamond NV centre with a single, well-defined value for the spatial location of the oscillator. It must thus be interpreted as representing a spin which is fully oriented along the $z$-axis, and an oscillator in a (classical or generalized coherent) state centered at position $x \pm d$.  The position of the oscillator can be registered and recorded, and thus effectively serves as a measurement of the spin state. Notice that the final state is once again a stable configuration which is insensitive to further non-unitary influences. As long as no sufficiently massive superpositions are created, its further time evolution will be unitary for all practical purposes, and its norm $N$ will be conserved.

We can repeat the above experiment many times, and each time record which of the two states in equation \eqref{final} is the outcome of that particular realization. The number of experiments resulting in a state with the oscillator centered at $x+d$ divided by the total number of conducted experiments will then be seen to converge to the value $|\alpha|^2$, as indicated in figure \ref{plotC}. Since it is impossible to predict which of the two outcomes will be realized in any one particular experiment, this is equivalent to having a probabilistic reduction of the initial state to just one of its components, with the probability for obtaining a given component set by its squared weight in the initial wavefunction:
\begin{align}
\left[ \alpha \chi_{\uparrow} + \beta \chi_{\downarrow} \right] \psi^0(x) \rightarrow \left\{ \begin{array}{lr} \chi_{\uparrow} \psi^0(x+d), & P = | \alpha |^2\phantom{.} \\ \chi_{\downarrow} \psi^0(x-d), & P = |\beta|^2. \end{array} \right.
\end{align}
Notice that for a general initial norm $|N_0|^2 \equiv |\alpha|^2 + |\beta|^2$, we would find probabilities $|\alpha^2| / |N_0|^2$ and $|\beta|^2 / |N_0|^2$ instead.

As noted before, the reduction process would happen unmeasurably fast if we were to employ any of the truly massive devices that are customarily used as quantum measurements machines. In that case, the dynamics described here thus reproduces the instantaneous `collapse of the wavefunction' that is traditionally postulated as an addition to Schr\"odinger's equation. For devices much lighter than the proposed micromechanical oscillator on the other hand, the non-unitary reduction process takes an unmeasurably long time to complete, and the dynamics remains effectively unitary, as prescribed by Schr\"odinger's equation. Both the unitary quantum dynamics of microscopic particles, and the non-unitary behavior of classical objects thus emerge naturally from the dynamics of equation \eqref{MSE}. In particular, the usual probabilistic predictions of quantum mechanics are recovered, in spite of the absence of a conserved norm of the wavefunction. 

\section{Appendix B: The Perturbation to Schr\"odinger's Equation}
The dynamical Schr\"odinger equation can be seen to be subject to spontaneous symmetry breaking, in the sense that its result may be qualitatively affected in the thermodynamic limit by only an infinitesimally weak unitarity breaking field. This description of broken unitarity is a straightforward extension of the well known mechanism of spontaneous symmetry breaking in crystals, magnets, and so on \cite{vWezel:Review}. For such spontaneous symmetry breaking to be effective in large, but finite, objects, there must exist a small, but finite, non-unitary contribution to its dynamics. In this appendix we give a brief sketch of how a non-unitary such as the one in \eqref{MSE} may be seen to naturally arise from considerations of the interplay between quantum mechanics and general relativity \cite{vWezel:Review}.

As it turns out, the fundamental building blocks of these theories (unitarity for quantum mechanics and general covariance for general relativity) are mutually incompatible concepts. All theories trying to bridge the gap between general relativity and quantum mechanics have to abandon either one or both of these principles. For example, the background dependence of non-equilibrium string theory breaks general covariance, while an equilibrium formulation in Euclidean spacetime is necessarily dissipative and thus non-unitary. The conflict can be made apparent by comparing simple examples of time evolution in the two theories \cite{Penrose}.  For the purpose of this paper we remain agnostic about the ultimate role of general covariance, but we assume that unitarity is not an exact symmetry of nature, and may thus be spontaneously broken \cite{vWezel:Review}. 

The density of the resonator in our experiment is rather modest compared to the densities occurring in black holes or in high energy physics. The dynamics of the resonator will therefore  be very close to purely quantum mechanical, and we expect to be able to include the effects of general relativity as minor perturbations to Schr\"odinger's equation. Rather than identifying the operator $d / d t$ with just the unitary $-(i / \hbar) \hat H$, we therefore introduce a small perturbative term $\hat{H}'$: 
\begin{align}
i \hbar \frac{d}{dt} \psi(\vec{r},t,\sigma) = \left[\hat{H} + \hat{H}'\right] \psi(\vec{r},t,\sigma),
\label{SE}
\end{align}
where the perturbation $\hat{H}'$ is assumed to be due to the influence of general relativity.

To find the functional form of the perturbation, we start by assuming that the defining property of the conflict between general relativity and quantum mechanics is the essential non-unitarity of the former. Other ingredients (such as a possible non-linearity) will be left to higher order terms. The correction term $\hat{H}'$ then cannot be a Hermitian operator, but on general grounds we would like the time evolution to be invertible even in the presence of gravity, so that physics makes sense if time flows backwards as well as forwards (although it does not necessarily have to look the same). One possible way to enforce this constraint is by writing
\begin{align}
\frac{d}{d t} \psi = - \frac{i}{\hbar} \left( \hat{H} - i \hat{X} \right) \psi,
\label{iX}
\end{align}
where $\hat{X}$ {\em is} a linear and Hermitian operator. 
The fact that the time evolution generated in this way does not conserve energy (as measured by $\hat{H}$) agrees with the lack of a locally conserved energy concept in a non-static configuration of general relativity. Of course globally, energy should be a conserved quantity, and we will have to choose $\hat{X}$ such that it takes account of that restriction. In fact, the presence of an order parameter in rigid macroscopic objects can be shown to automatically restore global energy conservation in the thermodynamic limit \cite{vWezel:Review}. 

\begin{figure}[t]
      \begin{center}
      \includegraphics[width=0.6\columnwidth]{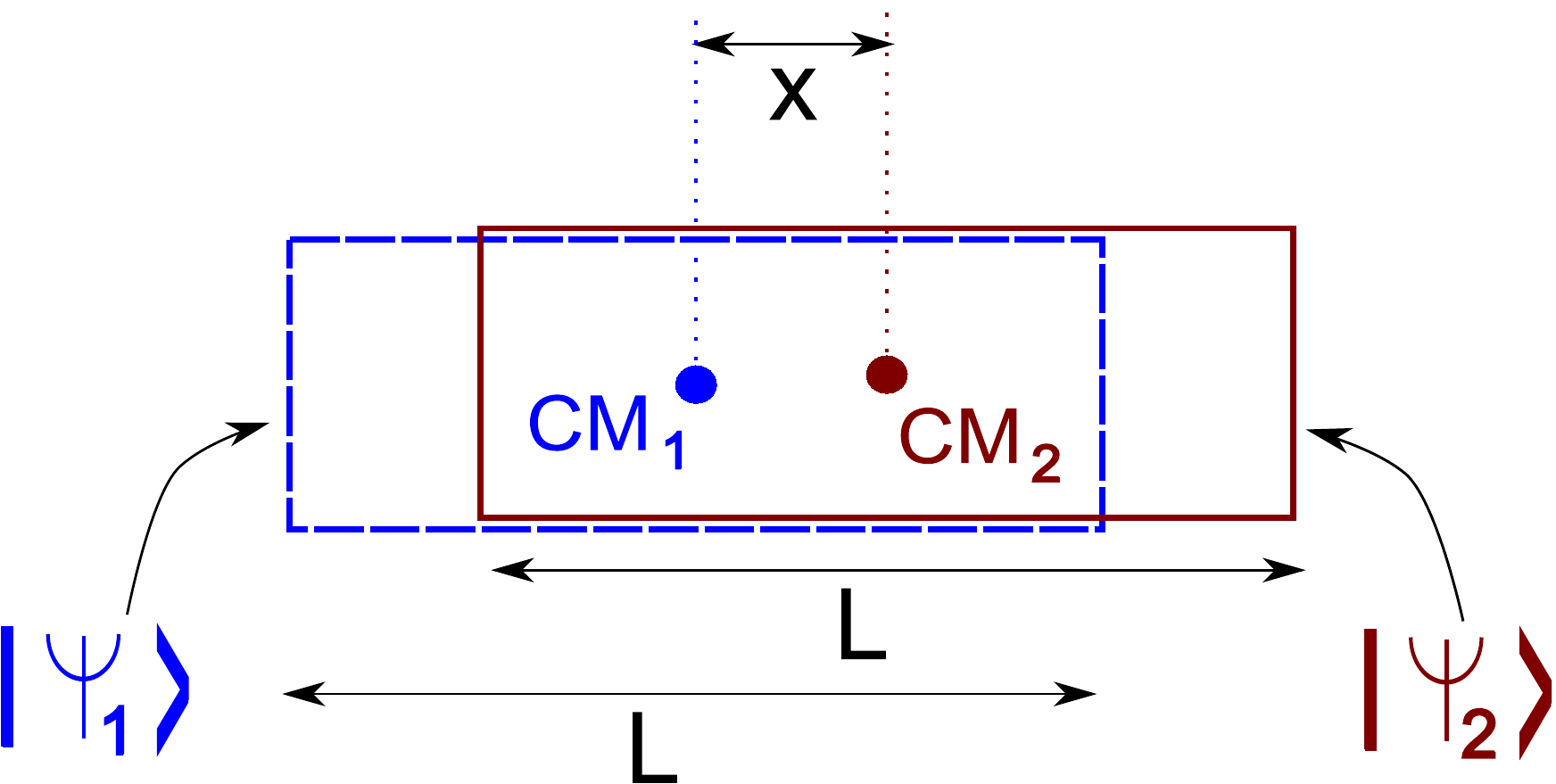}
      \end{center}
      \caption{The massive superposition of a block of width $L$ over a small distance $x$.}
      \label{fig5}
\end{figure}

The second step in our argumentation is to relate the `energy' term $\hat{X}$ to a measure for the extent to which the presence of a given quantum superposition is in conflict with the requirements of general relativity, and thus gives rise to a non-unitary correction to quantum mechanics. 
It has been shown by Di\'osi \cite{Diosi}, and independently by Penrose \cite{Penrose}, that there exists a covariant way of constructing such a measure. 
For an equal superposition of  two different mass distributions $\rho_1$ and $\rho_2$ in the Newtonian limit, they found this measure to be \cite{Diosi,Penrose}:
\begin{align}
\Delta = -4 \pi G \int \int \frac{\left[ \rho_1(x)-\rho_2(x) \right] \left[ \rho_1(y)-\rho_2(y) \right]}{\left| x-y \right|} \ d^3x \ d^3y.
\label{Newton2}
\end{align}
In the special case in which the superposed mass $m$ is a block which is evenly superposed over a distance $x$ small compared to its width $L$ (as in figure \ref{fig5}), the integrals of equation \eqref{Newton2} can be evaluated to yield \cite{vWezel:PhilMag}
\begin{align}
\Delta = G \frac{m^2}{2 L^3} x^2.
\label{Block}
\end{align}

The form of equation \eqref{Block} allows a straightforward generalization to the case of a generic superposition which consists of any number of components carrying arbitrary weights in the wavefunction. If $\hat{x}$ is the standard quantum mechanical operator measuring the position of the block's centre of mass (with the zero of position at the overall centre of mass of the initial wavefunction), then we can interpret $\Delta$ as the expectation value of the quantum operator
\begin{align}
\hat{\Delta} =  G \frac{m^2}{2 L^3} \hat{x}^2.
\label{DeltaOp}
\end{align}

Notice that although we started with a semi-classical definition of the incompatibility measure in equation \eqref{Newton2}, the final expression of equation \eqref{DeltaOp} is a fully quantum mechanical, Hermitian operator. This quantized form is only applicable to the specific case of a massive block superposed over small distances along one spatial direction, and in that sense is not as general as the original semiclassical expression. However, because it is an operator form it can be applied to any such wavefunction of the massive block, and it is not restricted to an even distribution over just two states. For such a more general wavefunction, the expectation value of the operator is still a good measure of the uncertainty introduced into the concept of a locally conserved energy by the application of general covariance. These properties of the operator form of $\hat{\Delta}$ suggest that it may also be an appropriate form for the first order correction to Schr\"odinger's time evolution in equation \eqref{iX}:
\begin{align}
\hat{X} \propto G \frac{m^2}{2 L^3} \hat{x}^2.
\label{Xop}
\end{align}

Finally, we need to address the fact that the spatial distance between points in separate space-times is an ill-defined notion in general relativity \cite{Penrose}. In this regard, the expression of equation \eqref{Xop} for the first order correction is not fully satisfactory. 
The operator $\hat{x}$ measures the position of the centre of mass in a component of the massive superposition using a universal coordinate system which is applied equally to all other components. We needed to define such a 'best possible' match between the coordinate systems of different components to first calculate the semi-classical measure for its inappropriateness (equation \eqref{Newton2}), but the final form of the correction to Schr\"odinger's equation should not depend on any such (arbitrary) specific choice of coordinate system. 
To model the ill-definedness in the definition of $\hat{x}$ we introduce a {\it random variable} $\xi$, and replace the measure of distance by $(\hat{x}-\xi)$.
\begin{align}
\hat{X} = G \frac{m^2}{2 L^3} \left[\hat{x}-\xi\right]^2
\label{Xfinal}
\end{align}
As emphasized also in the main text, the introduction of a random variable here should be seen only as a poor man's approach towards modeling an essentially ill-defined quantity. We do not expect the full theory of quantum gravity to be non-deterministic, but we are forced to take into account an {\it effectively} random correction to the notion of position if we are to consider gravity's first order perturbation to quantum dynamics. The equality of the numerical pre-factors of $\hat{x}$ and $\xi$ reflects the fact that the correction due to gravity should depend on the spatial spread of the wavefunction.

We thus finally reproduce equation \eqref{MSE} as the  effective, phenomenological description of the  minimal correction to the unitary Schr\"odinger equation due to the influence of general covariance:
\begin{align}
\frac{d}{d t} \psi = - \frac{i}{\hbar} \left( \hat{H} - i  G \frac{m^2}{2 L^3} \left[\hat{x}-\xi\right]^2 \right) \psi,
\end{align}

This final form of the modified Schr\"odinger equation can be straightforwardly integrated using standard numerical methods to yield a phenomenological prediction for the expected time evolution of the proposed resonator experiment.

\subsubsection{Acknowledgements}
The part of this work done at Argonne National Laboratory was supported by the US DOE, Office of Science, under Contract No. DE-AC02-06CH11357.

\end{document}